\documentclass[aps,prd,twocolumn,fleqn,superscriptaddress]{revtex4}
\usepackage{graphicx,color,natbib}
\usepackage{amsmath,amssymb,amsfonts}

\newcommand{\bse}{\begin{subequations}}
\newcommand{\ese}{\end{subequations}}
\newcommand{\be}{\begin{equation}}
\newcommand{\ee}{\end{equation}}
\newcommand{\bea}{\begin{eqnarray}}
\newcommand{\eea}{\end{eqnarray}}
\newcommand{\ba}{\begin{array}}
\newcommand{\ea}{\end{array}}
\newcommand{\h}{\frac{1}{2}}

\def\FF{{\mathcal{F}}}
\def\HH{{\mathcal{H}}}
\def\B{{\mathcal{B}}}
\def\ZZ{{\mathcal{Z}}}

\input amssym.def
\input amssym.tex

\usepackage[colorlinks=true, linkcolor=blue, bookmarks=true]{hyperref}

\begin{document}
IPM/P-2013/032

\title{Chiral Symmetry Breaking:\\ To Probe Anisotropy and Magnetic Field in QGP}

\author{Mohammad Ali-Akbari\footnote{aliakbari@theory.ipm.ac.ir}}
\affiliation{School of Particles and Accelerators, Institute for Research in Fundamental Sciences (IPM),
P.O.Box 19395-5531, Tehran, Iran}
\author{Hajar Ebrahim\footnote{hebrahim@ipm.ir}}
\affiliation{School of Physics, Institute for Research in Fundamental Sciences (IPM),
P.O.Box 19395-5531, Tehran, Iran}

\begin{abstract}
We discuss the (spontaneous) chiral symmetry breaking in a strongly coupled anisotropic quark-gluon plasma (QGP) in the presence of the magnetic field, using holography. The physical quantities related to the chiral symmetry breaking ($m, B_c$) distinguish between the effects of the anisotropy and magnetic field on the plasma. Anisotropy affects the system similar to the temperature and for its larger values heavier quarks can live in the QGP without getting condensed. Raising the anisotropy in the system will also increase the value of the critical magnetic field, $B_c$, at which the spontaneous chiral symmetry breaking happens. Both of these growths are even more when the magnetic field is applied perpendicular to the anisotropy direction. Such behaviour persists in the high temperature limit where the temperature is kept fixed. However, when the entropy density is held fixed, as one increases the anisotropy lighter mesons melt when the magnetic field is applied along the anisotropy direction, in contrast to when the magnetic field is perpendicular to the anisotropy direction.   
\end{abstract}

\maketitle


\section{Introduction}

Heavy ion collisions at RHIC and LHC produces a new phase of matter called Quark-Gluon Plasma (QGP). Hydrodynamic simulations of the QGP indicate that viscosity over entropy density is small and the plasma is strongly coupled \cite{Shuryak:2003xe}. Therefore the perturbation theory is not applicable. During the very early stages after the collision the plasma is formed and is out of equilibrium. Viscous hydrodynamic description applies after a certain time called $\tau_{iso}$ where there still exists a significantly different pressure anisotropy between longitudinal and transverse directions \cite{Chesler:2009cy}. Furthermore at this period of time simulations suggest that a strong magnetic field is produced \cite{Kharzeev:2007jp}. Therefore studying the characteristics of the QGP which can distinguish between the anisotropy and magnetic field is very interesting.

Gauge/Gravity duality is a good candidate to study strongly coupled field theories \cite{Maldacena}. It states that the classical gravity on asymptotically AdS background in $d+1$ dimensions is dual to $d$-dimensional strongly coupled Yang-Milles (YM) theory living on the boundary of AdS. Mateos, et al have extended this duality to spatially anisotropic finite-temperature background which asymptotically becomes $AdS_5\times S^5$ \cite{Mateos:2011tv}. It corresponds to an anisotropic ${\cal{N}}=4$ SYM plasma at non-zero temperature. Various aspects of this background has been studied in \cite{Rebhan:2012bw, Giataganas:2012zy}. Adding probe branes to this background provides us with the appropriate framework to add fundamental matter (quarks) to the YM theory \cite{Karch:2002sh} and study its properties in the presence of magnetic field.

The question we try to address in this paper is how differently the magnetic field and anisotropy affect the properties of the QGP. A good candidate is to study \textit{spontaneous} chiral symmetry breaking. In the gauge/gravity picture the chiral symmetry breaking in a strongly coupled plasma corresponds to the phase transition between the two different embeddings, Minkowski and black hole, of the probe brane \cite{Mateos:2006nu}. The two parameters used in this paper to discuss (spontaneous) chiral symmetry breaking are the mass and the (critical) magnetic field at which the condensation becomes non-zero. These two parameters are obtained from the asymptotic shape of the probe brane. In the anisotropic background introduced in \cite{Mateos:2011tv} we will embed a D7-brane and study the effect of anisotropy and magnetic field on its shape.  


\section{Anisotropic Background}
The background we are interested in is an anisotropic solution of
the type IIb supergravity equations of motion. This
solution in the string frame is given by \cite{Mateos:2011tv} %
\begin{eqnarray} \label{one}%
 ds^2&=&-\FF\B u^{-2}dt^2+u^{-2}(dx^2+dy^2)+\HH u^{-2}dz^2\nonumber \\
 &+&\FF^{-1}u^{-2} du^2+e^{\h\phi} d\Omega_5^2, \nonumber \\
 d\Omega_5^2&=&d\theta^2+\sin^2\theta d\Omega_3^2+\cos^2\theta d\varphi^2, \nonumber \\
 \chi&=&az,\ \ \phi=\phi(u),
\end{eqnarray} %
where $a$ is a constant. $\chi$ and $\phi$ are axion and dilaton fields,
respectively. 
$\HH$, $\FF$ and $\B$ depend only on the radial direction, $u$.
In terms of the dilaton field, they are %
\bse\label{three}\begin{align} %
\HH&=e^{-\phi},\\
 \label{ff} \FF&=  \frac{e^{-\frac{1}{2}\phi}\left[ a^2 e^{\frac{7}{2}\phi}(4u+u^2\phi')+16\phi'\right]}{4(\phi'+u\phi'')}\,,
 \\
 \frac{\B'}{\B}&=\frac{1}{24+10
 u\phi'}\left(24\phi'-9u\phi'^2+20u\phi''\right)\,,
\end{align}\ese %
where the dilaton field satisfies the following third-order equation %
\begin{eqnarray}\label{eomphi}%
 &&\frac{256 \phi ' \phi ''-16 \phi '^3
   \left(7 u \phi '+32\right)}{u \, a ^2 e^{\frac{7 \phi }{2}}
   \left(u \phi '+4\right)+16 \phi '} 
   +\frac{\phi ' }{u \left(5 u \phi '+12\right) \left(u \phi
   ''+\phi '\right)}\nonumber \\
   && \times \Big[13 u^3 \phi '^4+8 u \left(11
   u^2 \phi ''^2-60\phi''-12 u \phi ''' \right)\nonumber \\
   &&+ u^2 \phi '^3
   \left(13 u^2 \phi ''+96\right)  
    +2 u \phi '^2 \left(-5 u^3 \phi
   '''+53 u^2 \phi ''+36\right)\nonumber \\
   &&+ \phi ' \left(30 u^4 \phi
   ''^2-64 u^3 \phi '''-288+32 u^2 \phi
   ''\right) \Big]=0 \,.
\end{eqnarray}%
Note that the solution also contains a self dual five-form
field. 

The function $\FF(u)$ in the time and radial metric coefficients is the blackening factor. Therefore the horizon is located at $u=u_h$ where $\FF(u_h)=0$ and the
Hawking temperature is given by $T=-\frac{1}{4\pi}\FF'(u_h)\sqrt{\B(u_h)}$.
The boundary lies at $u=0$ and the
metric approaches $AdS_5\times S^5$
asymptotically. At the boundary, the suitable boundary conditions are %
\bse\begin{align}%
 \label{BC1}\phi_B&=0,\\
 \label{BC2}\FF_B&=\B_B=1.
\end{align}\ese %
The coordinates of the space-time where the gauge theory lives are $(t,x,y,z)$ where there is a $U(1)$ symmetry in the $xy$-plane. We call $x$ and $y$ the transverse directions and the
longitudinal direction is $z$. An anisotropy is clearly seen between
the transverse and longitudinal directions.

In the gauge theory side, the axion corresponds to a
position-dependent $\theta$-term or, more precisely, $\theta
\varpropto z$. The $\theta$-term is considered as an external source
which breaks the original isotropy of the system and forces the
system into an anisotropic equilibrium state. This external source
leads to a non-zero conformal anomaly meaning that the trace of the
energy-momentum tensor is no longer zero. In fact the trace of the
energy-momentum tensor is proportional to the anisotropy parameter,
$a$, which appears in the definition of the axion field as a
constant with the dimension of energy. In the gravity side this is
supported by the fact that diffeomorphism invariance in the radial
direction is broken in the process of the renormalization of the
on-shell action.

The broken conformal symmetry induces a new scale, $\mu$, in the
gauge theory. This scale appears in the thermodynamical quantities
and rescaling of this scale is realized as the freedom in the choice
of scheme. Therefore the gauge theory has three scales, $T$(which is
identified with the Hawking temperature of the background), $a$ and
$\mu$.

Since the equation of motion for the dilaton field is a third-order
differential equation, we need two initial conditions to solve it.
In order to specify these initial conditions,
it is better to define $\tilde{\phi}(u)\equiv\phi(u) +\frac{4}{7}\log a$.
This redefinition eliminates $a$ in \eqref{eomphi} and generates an
overall factor of $a^{2/7}$ in \eqref{ff}. Then one can expand the
dilaton field as %
\be\label{extend} %
 \tilde{\phi}=\tilde{\phi}_h+\sum_{n\geqslant1}\tilde{\phi}_n(u_h)(u-u_h)^n,
\ee %
where $\tilde{\phi}_h=\tilde{\phi}(u_h)$. Using \eqref{eomphi}, the expressions for the first two coefficients of \eqref{extend} have been introduced in \cite{Mateos:2011tv}. %
For given values of $u_h$ and $\tilde{\phi}_h$, the above initial
conditions and the boundary condition \eqref{BC1} help us solve the
equation of motion for $\phi$, numerically. Afterwards the other
components of the metric are found by using \eqref{three} and
\eqref{BC2}. Namely, the solution is characterized by two
parameters: the value of the dilaton field at the horizon and the
location of the horizon. In the case of $a=0$, the solution reduces to
an isotropic black D3-brane solution.


\subsection{High temperature limit} %
In the high temperature limit, $T\gg a,\mu$, the
solution has analytically been found. In this limit, up to the leading order in
$a$, the functions $\FF$, $\B$ and the dilaton field are given by %
\bse\begin{align}%
 \FF&=1-\frac{u^4}{u_h^4}+a^2\hat{\FF}_2(u), \\
 \B&=1-a^2\frac{u_h^2}{24}\left(\frac{10u^2}{u_h^2+u^2}+\log(1+\frac{u^2}{u_h^2})\right),\\
 \phi&=-a^2\frac{u_h^2}{4}\log(1+\frac{u^2}{u_h^2}),
\end{align}\ese
where %
\begin{eqnarray}%
 \hat{\FF}_2(u)&=&\frac{1}{24u_h^2}\bigg(8u^2(u_h^2-u^2)-10u^4\log2\nonumber \\
 &+&(3u_h^4+7u^4)\log(1+\frac{u^2}{u_h^2})\bigg),
\end{eqnarray}
The temperature and entropy density  of the solution in terms of the anisotropy parameter is %
\bse\begin{align} %
 \label{temp}T&=\frac{1}{\pi u_h}+\frac{(5\log2-2)u_h}{48\pi^2}a^2,\\
 \label{ent}\frac{s}{N_c^2}&=\frac{\pi^2 T^3}{2}+\frac{T}{16} a^2,
\end{align}\ese %
where $N_c$ is the number of colours.

It was verified in \cite{Mateos:2011tv} that there is a one-to-one map between $(\tilde{\phi},u_h)$ and $(a,T)$.
For the high temperature solution, this map can explicitly be written as %
\bse\begin{align} %
 u_h&=\frac{1}{\pi T}+\frac{5\log2-2}{48\pi^3T^3}a^2, \\
 \tilde{\phi}_h&=-\frac{a^2u_h^2}{4}\log2+\frac{4}{7}\log a.
\end{align}\ese
Therefore we can fix the temperature and discuss the behaviour of various physical quantities with respect to the anisotropy parameter. But in the general case such formulas can not be explicitly obtained. Thus we have to study the dependence of the physical quantities on $\frac{a}{T}$ for given $\tilde{\phi}_h$ and $u_h$.

\section{Fundamental matter in the anisotropic background}
In order to add the fundamental matter to the $SU(N)$ gauge theory we have to introduce a D7-brane into the anisotropic background in the probe limit. The probe limit means that the D7-brane does not modify the geometry. In fact the open strings stretched between probe D7-brane and the D3-D7 system leading to the geometry \eqref{one} give rise to the matter in the fundamental representation of the gauge group. In the large t' Hooft coupling and $N$ limits the dynamics of the open strings is described by the DBI action \cite{Myers:1999ps}
\be %
S_{DBI}=-\tau_7\int d^8\xi\  e^{-\phi} \sqrt{\det(G_{ab}+2\pi\alpha'F_{ab})}.
\ee %
The D7-brane tension is $\tau_7$ where $\tau_7^{-1}=(2\pi)^7 l_s^8g_s$  and $G_{ab}=g_{MN}\partial_a X^M\partial_b X^N$ where $g_{MN}$ is the background metric given by \eqref{one}. The D7-brane is extended along $t,x,y,z,u$ and wrapped around $S^3\subset S^5$. Although the four-form and the axion fields are non-zero in the background, in such an embedding the Chern-Simon action has no contribution to the dynamics of the D7-brane. We turn on the magnetic field on the probe brane along two different directions, $B_x=F_{yz}$ and $B_z=F_{xy}$, where $F_{ab}$ is the gauge field strength on the probe brane. The shape of the brane is given by the transverse directions $\theta$ and $\phi$ where we assume $\phi$ to be zero and $\theta$ depends on the radial direction. Therefore, in the presence of the magnetic field, the Lagrangian reduces to 
\bea\label{lagrangian} %
&&{\cal{L}}=e^{-\phi(u)}\frac{\cos^3\theta(u)}{u^5} \times\\
&& \sqrt{\B\ZZ^3\left(\HH+(2\pi\alpha')^2u^4(B_x^2+\HH B_z^2)\right)(1+u^2\FF \ZZ \theta'(u)^2)}.\nonumber
\eea%
The physical parameters we are interested to obtain can be found from the asymptotic solution to $\theta(u)$ equation of motion, $\theta(u)=\theta_0 u+\theta_2 u^3+ \dots$ \cite{Kruczenski:2003uq},
where $m=\frac{\theta_0}{2\pi\alpha'}$ is the mass of the fundamental matter and $c=\theta_2-\frac{1}{6}\theta_0^3$ corresponds to condensation that is proportional to $\langle \bar{\psi}\psi \rangle$. 

If we set the anisotropy parameter and the magnetic fields to zero the shape of the brane can be classified into two categories \cite{Mateos:2006nu}, one is the Minkowski embedding and the other one is the black hole embedding. Minkowski embedding means that the probe brane does not see the horizon and the quark and anti-quark bound states are stable (mesonic phase). Conversely, in the black hole embedding the probe brane crosses the horizon and the quark-antiquark bound states are unstable (melted phase). In fact a first order phase transition between these two embeddings may be identified with the chiral phase transition on the gauge theory side. Note also that there is another embedding called the critical embedding where the probe brane touches the horizon. 
\begin{figure}[ht]
\begin{center}
\includegraphics[width=2.6 in]{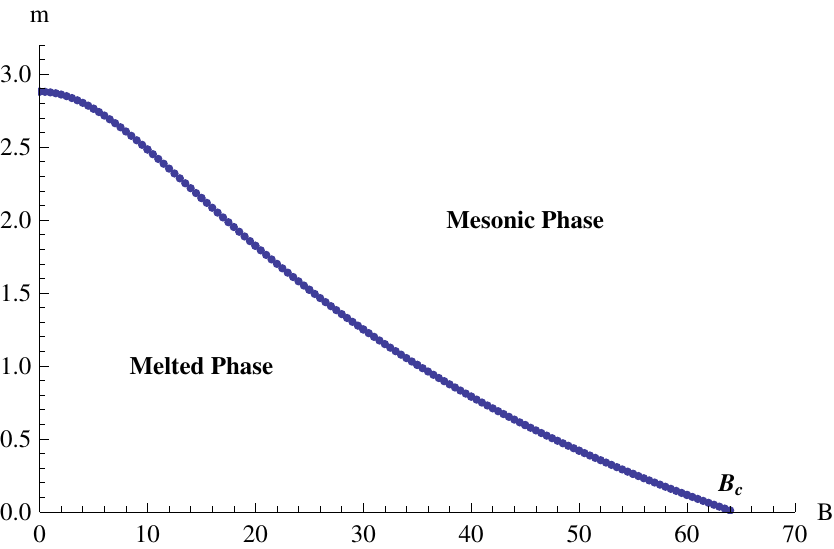}
\caption{ Critical embedding mass in terms of $B$ for $a=0$, $T=1$.}
\label{mB}
\end{center}
\end{figure}%

\begin{figure}[ht]
\begin{center}
\includegraphics[width=2.6 in]{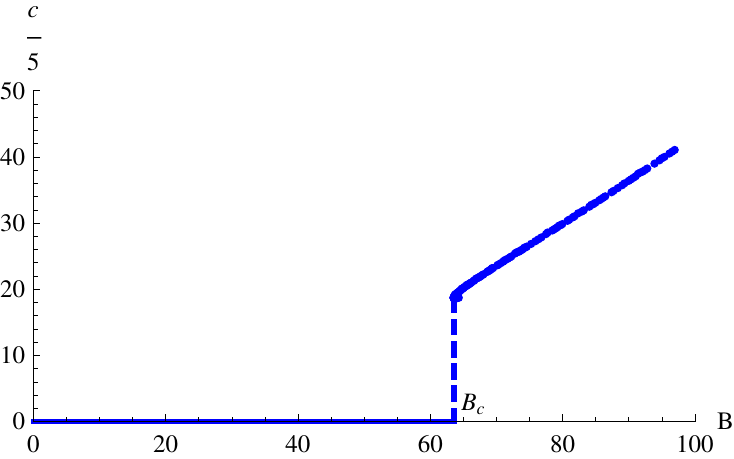}
\caption{ $c$ in terms of $B$ for $a=0$, $T=1$.}
\label{cB}
\end{center}
\end{figure}

Now we turn on the magnetic field but keep the anisotropy parameter zero, therefore $B_z$ and $B_x$ will affect the system similarly. We consider $B_x=B$ and $B_z=0$. Such system has been studied in \cite{Erdmenger:2007bn}. We solve the equation of motion for $\theta(u)$ numerically and the dependence of the mass on the magnetic field can be obtained from its asymptotic form. The result is shown in figure \ref{mB}. We see that for each value of the magnetic field there is a maximum value for the mass after which the system is at the mesonic phase. Note that for a certain value of the magnetic field, $B_c$, where the mass is zero, {\it{spontaneous}} chiral symmetry breaking happens. In figure \ref{mB}, $B_c\sim 64$. On the $m=0$ axis the chiral symmetry is spontaneously broken for $B$ larger than $B_c$. Thus the condensation, $c$, is non-zero although the mass is zero. On the contrary, $B< B_c$ corresponds to both $c$ and $m$ equal to zero. Therefore the chiral symmetry is restored. This has been shown in figure \ref{cB}. 

\subsection{High Temperature Limit}

\begin{figure}[ht]
\begin{center}
\includegraphics[width=2.6 in]{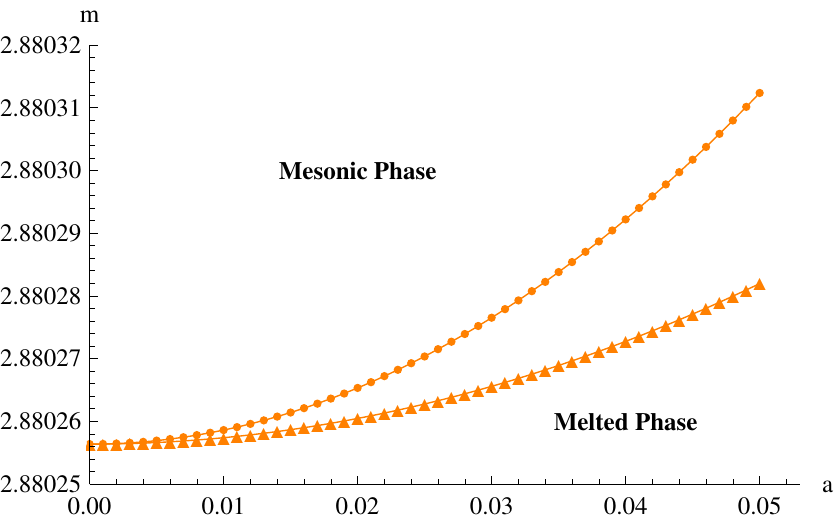}
\caption{Critical embedding mass in terms of $a$ for $B=0$ and $T=1~(\frac{s}{N_c^2}=\frac{\pi^2}{2})$, circles (triangles). The plots are fitted with $m= 2.88026 + 0.0223983 a^2~(2.88026 + 0.0102396 a^2)$.}
\label{mauh}
\end{center}
\end{figure}

Now we set the magnetic field to zero and switch on the anisotropy parameter. In the high temperature limit we choose $a\ll T$. The value of the maximum mass, at which the phase transition happens, increases as one raises $a$ for both $T$ or $s$ held constant. This is shown in figure \ref{mauh}. As it is expected from the metric components at high temperature limit, the dependence of mass on $a$ for any given constant value of the temperature or entropy density behaves as $a^2$. For example for $T=1~(\frac{s}{N_c^2}=\frac{\pi^2}{2})$ the circular(triangular) points in figure \ref{mauh} are fitted with $m= 2.88026 + 0.0223983 a^2~(2.88026 + 0.0102396 a^2)$. Both curves coincide at $a=0$ since $T=1$ corresponds to $\frac{s}{N_c^2}=\frac{\pi^2}{2}$. Raise in the temperature will increase the mass. Above (below) each curve corresponds to mesonic (melted) phase of the fundamental matter. One can see from figure \ref{mauh} that compared to $a=0$, for larger values of $a$ heavier quarks can live in the QGP without getting condensed. This somehow indicates that anisotropy parameter behaves similarly to the temperature. 

According to \eqref{ent} as one raises $a$ the temperature decreases when the entropy density is kept fixed. Therefore, the fall in the temperature opposes the effect of the anisotropy parameter. Consequently, $m$ for fixed $s$ is less than its value for fixed $T$, at a given $a$, as it is observed in figure \ref{mauh}. 

\begin{figure}[ht]
\begin{center}
\includegraphics[width=2.6 in]{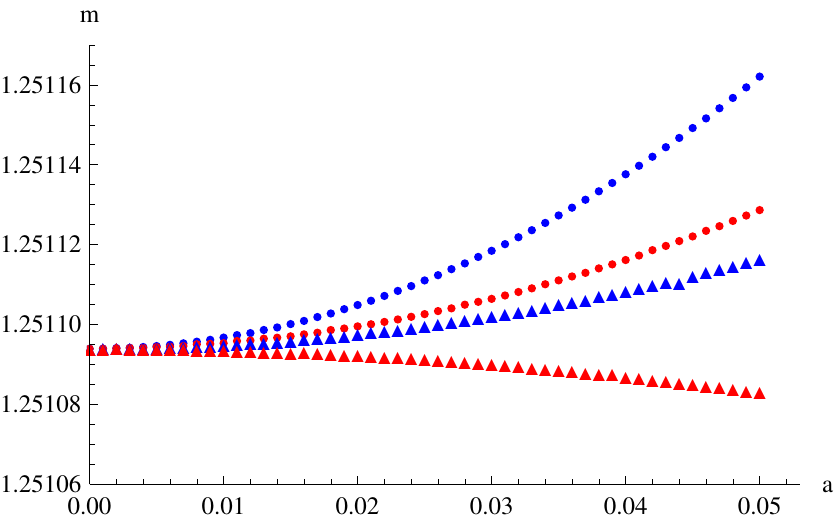}
\caption{ Critical embedding mass in terms of $a$ for $T=1~(\frac{s}{N_c^2}=\frac{\pi^2}{2})$, circles (triangles) and $B=30$. For the blue(red) graphs $B$ is along $x(z)$-direction.}
\label{maB}
\end{center}
\end{figure}

Intriguing observations can be made when both the magnetic field and anisotropy parameter are non-zero. We assume the cases where the magnetic field is along the anisotropic direction $B_z=B$ or it is perpendicular to the anisotropy direction $B_x=B$. Interestingly the behaviour of the mass is slightly different between these cases as has been shown in figure \ref{maB}. Note that the curves with circular(triangular) points represent fixed $T(\frac{s}{N_c^2})$. Compared to when $B_z\neq 0$, for $B_x\neq 0$ which is the magnetic field applied perpendicular to the anisotropy direction, the phase transition happens at larger values of mass. Therefore when the magnetic field is non-zero the phase transition between melted and mesonic phases will realise the presence of anisotropy in the system. In both cases of non-zero magnetic field the graphs can be fitted by functions as $m_{x(z)}=m_0+\alpha_{x(z)} a^2$ where by $m_{x(z)}$ we mean having $B_{x(z)}$ non-zero. $m_0$ and $\alpha_{x(z)}$ are constants depending on the value of the magnetic field. Therefore the difference between the masses where $B_{x}$ or $B_{z}$ is non-zero ($\delta m=m_x-m_z$) will be proportional to $a^2$.   

Although the anisotropy in the background does not change the tension of the probe brane, it is well known that the tension is effectively increased in the presence of the magnetic field \footnote{Consider a D7-brane in flat background. In the
presence of the magnetic field, the action becomes
$
S_{DBI}=-\tau_7\sqrt{1+B^2}\int d^{8}\xi=-\tau_{eff} \int d^{8}\xi,
$
where $B=F_{xy}$. All the other fields have been turned off. It is clearly seen that $\tau_{eff}\geq \tau_7$.}. The larger the magnitude of magnetic field, the larger the tension of the probe brane. As it is clearly seen from \eqref{lagrangian}, the magnetic field along the anisotropy direction is multiplied by $\HH(u)$. Since this function is equal to or larger than one for all values of the radial coordinate, the magnetic field along the anisotropy direction seems effectively larger. In other words, the anisotropy may result in an effective magnetic field, $B_{eff}$, which is bigger than $B_z$. Therefore, for equal values of the magnetic fields, $B_z=B_x=B$, the tension is larger in the longitudinal direction than transverse directions. As a result, when the magnetic field is turned on along the anisotropy direction the D-brane resists more against deformation and therefore the valve of the mass is less than the case where the magnetic field is along $x$, as observed in figure \ref{maB}. Moreover it indicates that the difference between these two tensions is more recognisable for larger values of the anisotropy parameter as our numerical results approve it. This argument can be applied in both cases where temperature or entropy density is held fixed. 

Following the discussion in the previous paragraph, since the blackening factor, ${\cal{F}}$, is multiplied by the magnetic field in \eqref{lagrangian} the effect of the temperature will be enhanced when the magnetic field is along the anisotropy direction, $B_z$, compared to $B_x$. Therefore using the argument describing the plots in figure \ref{mauh} the behaviour of the lowest curve in \ref{maB} may be explained. 
\begin{figure}[ht]
\begin{center}
\includegraphics[width=2.6 in]{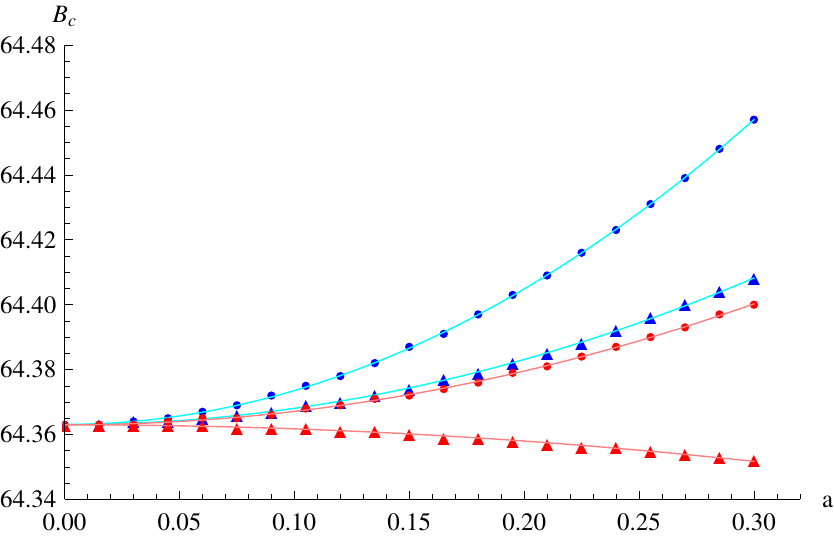}
\caption{$B_c$ in terms of $a$ for $T=1~(\frac{s}{N_c^2}=\frac{\pi^2}{2})$, circles (triangles). In the blue(red) ones, $B$ is along $x(z)$-direction. The graphs are all fitted with $B_c=B_{c0}+\beta a^2$ where $B_{c0}$ and $\beta$ are constant. \label{Bca}}
\end{center}
\end{figure}%

How the chiral symmetry breaking knows of the anisotropy in the system can also be seen in figure \ref{Bca}. It shows the dependence of the value of the critical magnetic field, $B_c$ at which the spontaneous chiral symmetry breaking happens, with respect to $a$. Apart from the nonzero $B_z$ at fixed entropy case, we observe that the chiral symmetry breaks spontaneously at higher values of the magnetic field as one increases $a$. We can also conclude that the critical value of the magnetic field is larger for $B_x\neq 0$ than $B_z\neq 0$ when $a$ is the same for both of them. Regarding the fact that $m_{x}$ is larger than $m_z$ in figure \ref{maB}, it is essential to have stronger magnetic field to spontaneously break the chiral symmetry. This also explains the observation in figure \ref{Bca}. 


In summary an appealing result from figures \ref{maB} and \ref{Bca} is that for a particular value of the anisotropy parameter the chiral symmetry is restored at larger values of the magnetic field if the magnetic field is turned on in the direction perpendicular to the anisotropy direction than if it is turned on along the anisotropy direction.

%

\subsection{General Values of The Anisotropy Parameter}
\begin{figure}[ht]
\begin{center}
\includegraphics[width=2.6 in]{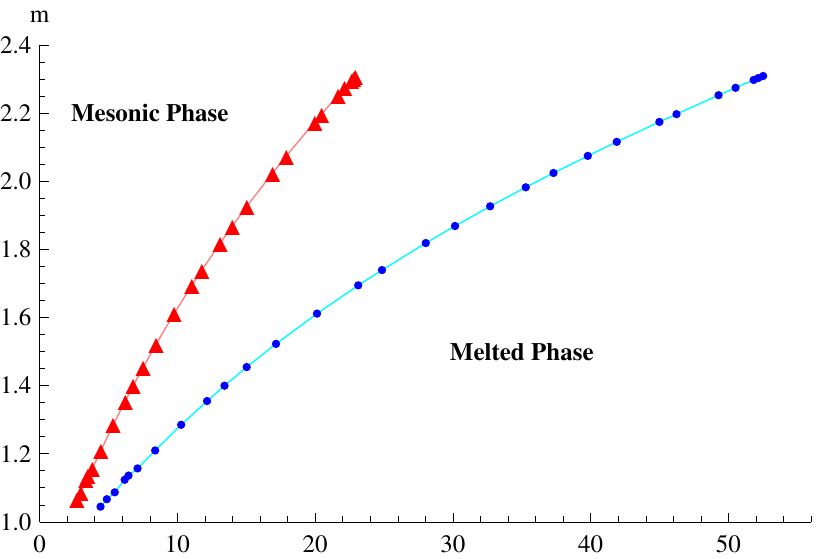}
\caption{Critical embedding mass in terms of $\frac{a}{T}$ (circles) and $a s^{\frac{-1}{3}}$ (triangles) for $B=0$, $u_h=1$.}
\label{generalma}
\end{center}
\end{figure}

In this section we will generalize the previous calculations for arbitrary values of the anisotropy parameter. For example analogous plot to \ref{mauh} for general $a$ is shown in figure \ref{generalma}. We will again see that for zero magnetic field if we increase the anisotropy parameter the mass of the critical embedding increases. Similar to figure \ref{mB} that the points divide the $m$ and $B$ phase space into two phases, melted and mesonic, we observe that the points in figure \ref{generalma} divide the $m$ and $\frac{a}{T} (a s^{\frac{-1}{3}})$ phase space into the same subspaces. For a given value of $m$ at $a=0$ and a fixed temperature or entropy density, where the system is at mesonic phase, by raising sufficiently the anisotropy the system will fall into the melted phase. Therefore the anisotropy parameter acts similarly to the temperature. Such behaviour is in contrast to the magnetic field where at a given value of $m$ for large enough values of the magnetic field the system is in the mesonic phase.
\begin{figure}[ht]
\begin{center}
\includegraphics[width=2.6 in]{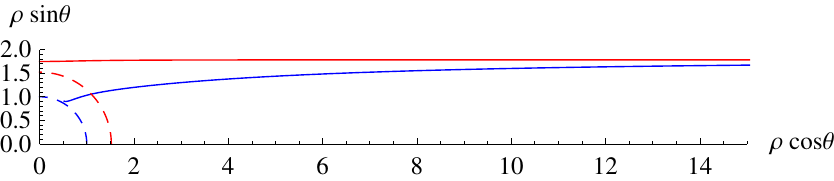}
\caption{ Shape of the brane for Minkowski(red) and black hole(blue) embeddings. In Minkowski embedding $a=0$ and $T=0.372871$ which corresponds to $u_h=0.853444$. In black hole embedding $a=5.60665$ and $T=0.372871$ which corresponds to choosing ${\tilde{\phi}}_h=0.22$ and $u_h=1$. Both shapes give the mass equal to 1.26022.}
\label{shape}
\end{center}
\end{figure}%
In order to explain this result more explicitly one can look at the shape of the brane, shown in figure \ref{shape}. Let us consider a Minkowski embedding in the black hole background with $u_h=0.853444$ and zero anisotropy parameter. Then we turn on the anisotropy parameter. Thus the shape of the brane with the same mass and temperature is described by a black hole embedding. Note that since the temperature is kept fixed the horizon for the black hole embedding lies at $u_h=1$. We have shown the horizon for each embedding with the corresponding colour \footnote{In order to define the transverse direction to the probe brane we have used the following change of coordinate
$\rho^2=\frac{1}{2}(\frac{1}{u^2}+\sqrt{\frac{1}{u^4}-\frac{1}{u_h^4}})$. In this coordinate the boundary of space-time is at $\rho\rightarrow \infty$.}. This means that by applying anisotropy the quark-antiquark bound state becomes unstable and the system goes into the melted phase. It shows that anisotropy is responsible for the dissociation. Our observation is consistent with the conclusion reported in \cite{Giataganas:2012zy}.

\begin{figure}[ht]
\begin{center}
\includegraphics[width=2.6 in]{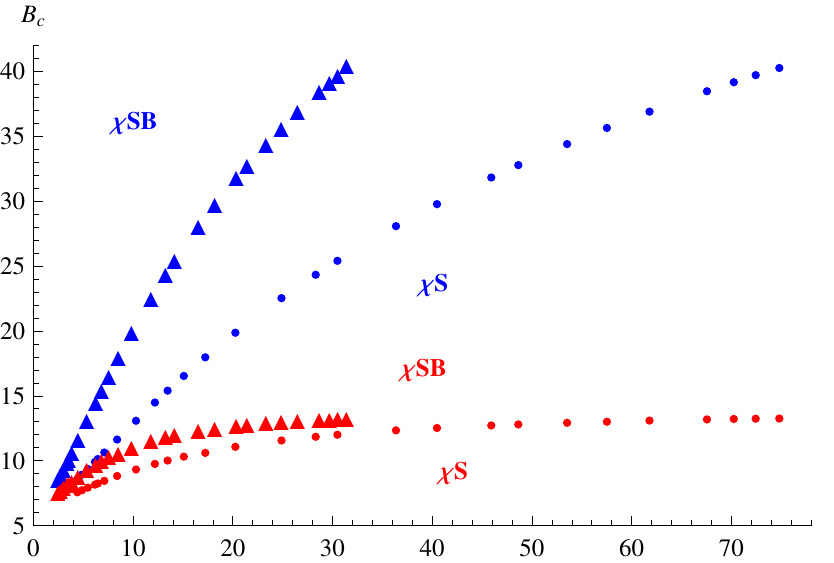}
\caption{$B_c$ in terms of $\frac{a}{T}$ (circles) and $a s^{\frac{-1}{3}}$ (triangles) for $u_h=1$. For the blue(red) points $B$ is along $x(z)$-direction.}
\label{generalBca}
\end{center}
\end{figure}

The dependence of the critical magnetic field on $\frac{a}{T} (a s^{\frac{-1}{3}})$ is plotted in figure \ref{generalBca}. As one increases the parameter $\frac{a}{T} (a s^{\frac{-1}{3}})$ the spontaneous chiral symmetry breaking happens at larger values of the magnetic field. Furthermore this growth in the value of $B_c$ is bigger if the magnetic field is applied perpendicular to the anisotropy direction. For the larger values of $\frac{a}{T} (a s^{\frac{-1}{3}})$, the difference between $B_x$ and $B_z$ becomes more recognizable. Therefore the value of the critical magnetic field might be a good characteristic of the anisotropy in the system. 

{\em Acknowledgement:} The authors have enjoyed participating in ``7th Crete Regional Meeting in String Theory'' where the idea of this paper was formed.



\begin{thebibliography}{99}

\bibitem{Shuryak:2003xe}
  E.~Shuryak,
  Prog.\ Part.\ Nucl.\ Phys.\  {\bf 53}, 273 (2004)
  \href{http://arxiv.org/abs/hep-ph/0312227}{[arXiv:hep-ph/0312227]};
  E.~V.~Shuryak,
  Nucl.\ Phys.\  A {\bf 750}, 64 (2005)
  \href{http://arxiv.org/abs/hep-ph/0405066}{[arXiv:hep-ph/0405066]}.
  
\bibitem{Chesler:2009cy}
  P.~M.~Chesler and L.~G.~Yaffe,
  Phys.\ Rev.\ D {\bf 82}, 026006 (2010)
   \href{http://arxiv.org/abs/0906.4426}{[arXiv:0906.4426 [hep-th]]};
  M.~P.~Heller, R.~A.~Janik and P.~Witaszczyk,
  Phys.\ Rev.\ Lett.\  {\bf 108}, 201602 (2012)
  \href{http://arxiv.org/abs/1103.3452}{[arXiv:1103.3452 [hep-th]]}.
  
\bibitem{Kharzeev:2007jp}
  D.~E.~Kharzeev, L.~D.~McLerran and H.~J.~Warringa,
  Nucl.\ Phys.\ A {\bf 803}, 227 (2008)
   \href{http://arxiv.org/abs/0711.0950}{[arXiv:0711.0950 [hep-ph]]}.
  
   \bibitem{Maldacena}
  J.~M.~Maldacena,
  Adv.\ Theor.\ Math.\ Phys.\  {\bf 2} (1998) 231
  [Int.\ J.\ Theor.\ Phys.\  {\bf 38} (1999) 1113]
  \href{http://arxiv.org/abs/hep-th/9711200}{[arXiv:hep-th/9711200]};
  S.~S.~Gubser, I.~R.~Klebanov and A.~M.~Polyakov,
  Phys.\ Lett.\  B {\bf 428} (1998) 105
  \href{http://arxiv.org/abs/hep-th/9802109}{[arXiv:hep-th/9802109]};
  E.~Witten,
  Adv.\ Theor.\ Math.\ Phys.\  {\bf 2} (1998) 253
  \href{http://arxiv.org/abs/hep-th/9802150}{[arXiv:hep-th/9802150]}.
  
\bibitem{Mateos:2011tv} 
  D.~Mateos and D.~Trancanelli,
  JHEP {\bf 1107}, 054 (2011)
  \href{http://arxiv.org/abs/1106.1637}{[arXiv:1106.1637 [hep-th]]}.

\bibitem{Rebhan:2012bw} 
  A.~Rebhan and D.~Steineder,
  JHEP {\bf 1208}, 020 (2012)
  \href{http://arxiv.org/abs/1205.4684}{[arXiv:1205.4684 [hep-th]]};
  A.~Rebhan and D.~Steineder,
  Phys.\ Rev.\ Lett.\  {\bf 108}, 021601 (2012)
  \href{http://arxiv.org/abs/1110.6825}{[arXiv:1110.6825 [hep-th]]};
  D.~Giataganas,
  PoS Corfu {\bf 2012}, 122 (2013)
  \href{http://arxiv.org/abs/1306.1404}{[arXiv:1306.1404 [hep-th]]};
  M.~Chernicoff, D.~Fernandez, D.~Mateos and D.~Trancanelli,
  JHEP {\bf 1208}, 100 (2012)
  \href{http://arxiv.org/abs/1202.3696}{[arXiv:1202.3696 [hep-th]]};
  L.~Patino and D.~Trancanelli,
  JHEP {\bf 1302}, 154 (2013)
  \href{http://arxiv.org/abs/1211.2199}{[arXiv:1211.2199 [hep-th]]};
  M.~Chernicoff, D.~Fernandez, D.~Mateos and D.~Trancanelli,
  JHEP {\bf 1208}, 041 (2012)
  \href{http://arxiv.org/abs/1203.0561}{[arXiv:1203.0561 [hep-th]]};
  K.~B.~Fadafan, D.~Giataganas and H.~Soltanpanahi,
  \href{http://arxiv.org/abs/1306.2929}{[arXiv:1306.2929 [hep-th]]};
  K.~B.~Fadafan and H.~Soltanpanahi,
  JHEP {\bf 1210}, 085 (2012)
  \href{http://arxiv.org/abs/1206.2271}{[arXiv:1206.2271 [hep-th]]};
  S.~-Y.~Wu and D.~-L.~Yang,
  JHEP {\bf 1308}, 032 (2013)
  \href{http://arxiv.org/abs/1305.5509}{[arXiv:1305.5509 [hep-th]]};
  I.~Gahramanov, T.~Kalaydzhyan and I.~Kirsch,
  Phys.\ Rev.\ D {\bf 85}, 126013 (2012)
  \href{http://arxiv.org/abs/1203.4259}{[arXiv:1203.4259 [hep-th]]}; 
  S.~Chakraborty and N.~Haque,
  Nucl.\ Phys.\ B {\bf 874}, 821 (2013)
  \href{http://arxiv.org/abs/1212.2769}{[arXiv:1212.2769 [hep-th]]}; 
  K.~A.~Mamo,
  JHEP {\bf 1210}, 070 (2012)
  \href{http://arxiv.org/abs/1205.1797}{[arXiv:1205.1797 [hep-th]]}.
  
\bibitem{Giataganas:2012zy} 
  D.~Giataganas,
  JHEP {\bf 1207}, 031 (2012)
   \href{http://arxiv.org/abs/1202.4436}{[arXiv:1202.4436 [hep-th]]}.

  
\bibitem{Karch:2002sh}
  A.~Karch and E.~Katz,
  JHEP {\bf 0206}, 043 (2002)
   \href{http://arxiv.org/abs/hep-th/0205236}{[hep-th/0205236]}.
  
\bibitem{Mateos:2006nu} 
  D.~Mateos, R.~C.~Myers and R.~M.~Thomson,
  Phys.\ Rev.\ Lett.\  {\bf 97}, 091601 (2006)
  \href{http://arxiv.org/abs/hep-th/0605046}{[hep-th/0605046]};
  D.~Mateos, R.~C.~Myers and R.~M.~Thomson,
  JHEP {\bf 0705}, 067 (2007)
  \href{http://arxiv.org/abs/hep-th/0701132}{[hep-th/0701132]}.
  

\bibitem{Myers:1999ps} 
  R.~C.~Myers,
  JHEP {\bf 9912}, 022 (1999)
  \href{http://arxiv.org/abs/hep-th/9910053}{[hep-th/9910053]}.

\bibitem{Kruczenski:2003uq}
  M.~Kruczenski, D.~Mateos, R.~C.~Myers and D.~J.~Winters,
  JHEP {\bf 0405}, 041 (2004)
  \href{http://arxiv.org/abs/hep-th/0311270}{[hep-th/0311270]};
  A.~Karch, A.~O'Bannon and K.~Skenderis,
  JHEP {\bf 0604}, 015 (2006)
  \href{http://arxiv.org/abs/hep-th/0512125}{[hep-th/0512125]};
  S.~Kobayashi, D.~Mateos, S.~Matsuura, R.~C.~Myers and R.~M.~Thomson,
  JHEP {\bf 0702}, 016 (2007)
  \href{http://arxiv.org/abs/hep-th/0611099}{[hep-th/0611099]};
  C.~Hoyos, T.~Nishioka and A.~O'Bannon,
  JHEP {\bf 1110}, 084 (2011)
  \href{http://arxiv.org/abs/1106.4030}{[arXiv:1106.4030 [hep-th]]}.

\bibitem{Erdmenger:2007bn}
  J.~Erdmenger, R.~Meyer and J.~P.~Shock,
  JHEP {\bf 0712}, 091 (2007)
   \href{http://arxiv.org/abs/0709.1551}{[arXiv:0709.1551 [hep-th]]};
  V.~G.~Filev, C.~V.~Johnson, R.~C.~Rashkov and K.~S.~Viswanathan,
  JHEP {\bf 0710}, 019 (2007)
   \href{http://arxiv.org/abs/hep-th/0701001}{[hep-th/0701001]};
  M.~S.~Alam, V.~S.~Kaplunovsky and A.~Kundu,
  JHEP {\bf 1204}, 111 (2012)
  \href{http://arxiv.org/abs/1202.3488}{[arXiv:1202.3488 [hep-th]]}.

\bibitem{Filev2}
  V.~G.~Filev,
  JHEP {\bf 0804}, 088 (2008)
   \href{http://arxiv.org/abs/0706.3811}{[arXiv:0706.3811 [hep-th]]}.
  
 
  

\end{thebibliography}
\end{document}